# Quantum Transport and Field-Induced Insulating States in Bilayer Graphene *pnp* Junctions

*Lei Jing[†], Jairo Velasco Jr.[†], Philip Kratz, Gang Liu[§], Wenzhong Bao, Marc Bockrath,*

*Chun Ning Lau\**

Department of Physics, University of California, Riverside, CA 92521

[§]Current Address: Department of Chemistry, University of California, Los Angeles, CA 90095

[†] These authors contribute equally to this work.

\* Email: lau@physics.ucr.edu

ABSTRACT

We perform transport measurements in high quality bilayer graphene *pnp* junctions with suspended top gates. At a magnetic field *B=0*, we demonstrate band gap opening by an applied perpendicular electric field, with an On/Off ratio up to 20,000 at 260mK. Within the band gap, the conductance decreases exponentially by 3 orders of magnitude with increasing electric field, and can be accounted for by variable range hopping with a gate-tunable density of states, effective mass, and localization length. At large *B*, we observe quantum Hall conductance with fractional values, which arise from equilibration of edge states between differentially-doped regions, and the presence of an insulating state at filling factor $\nu$=0. Our work underscores the importance of bilayer graphene for both fundamental interest and technological applications.

Bilayer graphene (BLG) is a unique two dimensional (2D) system with an unusual band structure -- parabolic bands with the conduction and valence bands touching at a point[1-3]. Consequently, its charge carriers behave as massive Dirac fermions, and are described by a combination of the Schrödinger and Dirac equations. This unique band structure gives rise to a number of novel properties, such as a gate tunable band gap[2], excitonic condensation[4, 5], and potentially novel integer and fractional quantum Hall (QH) states[6, 7 8]. On the technological front, band gap opening in bilayer graphene[2, 9, 10 11-14] in an applied electric field has generated significant interest as a promising route to band gap engineering and control in graphene electronics.

Here we report experimental investigation of dual-gated bilayer graphene junctions, which have mobility up to 10,000 cm$^2$/Vs, in zero and high magnetic fields. Using a combination of a Si back gate and a suspended top gate, these dual-gated structures offer *in situ* control of the dopant density $n$ and type of different regions, as well as independent tuning of $n_2$ and applied electric field $E_2$ for the region under the top gate. As $E_2$ is increased from 0 to 0.93 V/nm, the device conductance at charge neutrality point decreases exponentially by over 3 orders of magnitude, reaching an On/Off ratio of 20,000, that is more than 10 times higher than previously reported values[12, 13]. The data can be satisfactorily accounted for by variable range hopping (VRH) in 2D, suggesting the opening of an electric field induced band gap which enables the localization length, effective mass, and density of states to be tuned. In a high magnetic field $B$, we observe fractional-valued QH plateaus, which arise from edge state equilibration at the interface of differentially doped regions, in agreement with theoretical predictions[15, 16]. Notably, an insulating state develops at filling factor ν=0, whose conductance are exponentially dependent on applied *B*. Thus, our work suggests that competing symmetries and insulating states in bilayer graphene can be tuned by electric and magnetic fields, which are of significant interest for both technological applications and fundamental understanding of 2D systems.

BLG sheets are exfoliated from bulk graphene onto Si/SiO$_2$ wafers cleaned with a H$_2$SO$_4$/H$_2$O$_2$ solution mixture. The Ti/Al electrodes and Ti suspended top gates[17, 18] are fabricated by electron beam lithography. A device schematic is shown in Fig. 1a inset. Unless specified otherwise, the devices are measured at 260 mK in a He$^3$ refrigerator using standard lock in techniques. Here we focus on data from a single device with width *W*=1.2 µm, and source-drain separation *L*= 2.3 µm. The top gate, straddling the center of the device, is 550 nm long and suspended at *d*~50 nm above the substrate.

We first examine the device behavior as a function of back gate voltage $V_{bg}$. Fig. 1a plots the differential conductance *G* of the device *vs.* $V_{bg}$ at *B*=0, with the Ferimi level tuned to the Dirac point at $V_{bg}^0$≈-18V. The electron mobility is ~10,000 cm$^2$/Vs, while the hole mobility is significantly lower. Thus in the rest of the Letter we will focus on the electron-doped regime. The device conductance in units of $e^2/h$ at *B*=8T is shown in Fig. 1b. Theoretically, in a high magnetic field *B*, we expect the device's Hall conductance to be quantized at[1, 3, 10]

$$\sigma_{xy} = 4\nu \frac{e^2}{h}, \nu\text{=-3, -2, -1, 1, 2, 3....} \qquad (1)$$

where $\nu = \frac{nh}{Be}$ is the filling factor, *e* is the electron charge, *h* Planck's constant, and *n* the charge density. The omission of the *ν*=0 state arises from the degeneracy of the zeroth and first Landau levels (LL) and the resultant eight-fold degeneracy at zero energy. In Fig. 1b, clear plateaus that are quantized at *G*=8, 12, 16… $e^2/h$ are observed, in agreement with Eq. (1). Fig. 1c shows the standard LL "fan diagram", *i.e.* the evolution of the conductance plateaus with *B* and $V_{bg}$.

Impressively, a total of 15 plateaus are visible for $B>2T$, demonstrating the high quality of this device. Furthermore, since the trajectory of the center of a plateau has a slope $ve/h\alpha_{bg}$ in the $V_{bg}$-$B$ plane, where $\alpha_{bg}=n/V_{bg}$ is the back gate coupling efficiency, we extract $\alpha_{bg}\approx 7.4\times 10^{10}$ cm$^{-2}$V$^{-1}$ from Fig. 1b.

By applying voltages to both top and back gates, we can create *pnp* junctions with *in situ* modulation of junction polarity and dopant levels. Fig. 2a shows the conductance plot of $G$ (color) *vs.* $V_{bg}$ (vertical axis) and $V_{tg}$ (horizontal axis) at $B=0$. The plot can be partitioned into 4 regions with different combinations of dopant types, with the Dirac points occurring at $(V_{tg}^0, V_{bg}^0)=(-2.6V, -18V)$. In particular, the blue diagonal features correspond to the conductance of the region that is controlled by both $V_{tg}$ and $V_{bg}$, *i.e.*, the area under the top gate. From the slope of the diagonal features near the Dirac points, which corresponds to the ratio of the coupling efficiencies between the top gate and the back gate, we extract the top gate coupling efficiency $\alpha_{tg}\approx 1.2\times 10^{11}$ cm$^{-2}$V$^{-1}$. At high $V_{tg}$ and $V_{bg}$ values, the dark blue regions exhibit noticeable curvature, which arises from the deflection of the top gate under the electrostatic pressure.

This *in situ* creation of *pnp* or *npn* junctions in double gated junctions has been extensively studied in single layer graphene (SLG) devices[16, 17, 19-23], and enabled the observation of phenomena such as Klein tunneling[24-26], equilibration of counter-propagating edge modes[16, 17, 20, 21], and conductance fluctuations induced by charge localization in the quantum Hall regime[21]. For BLG, a unique aspect is the possibility for independent control of the charge density $n_2$ and the electric field $E_2$ applied across the bilayer[2, 9, 11-14, 27]. Here the subscript 2 denotes the top-gated region. Quantitatively, $V_{bg}$ induces an electric field below graphene $E_b=\varepsilon_{SiO}(V_{bg}-V_{tg}^0)/t$, where $\varepsilon_{SiO}\approx 3.9$ and $t=300$nm are the dielectric constant and thickness of the SiO$_2$ layer, respectively. Similarly, $V_{tg}$ induces $E_t=-(V_{tg}-V_{tg}^0)/d$ above graphene. The difference of the fields yields the total charge density under the top gate, $n_2=(E_b-E_t)e$, whereas their average $E_2=(E_b+E_t)/2$ breaks the inversion symmetry and yields a potential difference $V_2= E_2$ (3.3 Å) across the bilayer. For the data shown in Fig. 2a, $E_2$ ranges from 0 to 0.96 V/nm.

To better explore this unique aspect of BLG, we replot part of the data shown in Fig. 2a in terms of $n_2$ (horizontal axis) and $E_2$ (vertical axis), as shown in Fig. 2b (note the logarithmic color scale that spans 6 orders of magnitude.) To account for the deflection of the top gate under applied voltages, we self-consistently solve for the deflection and $E_t$ by considering the electrostatically induced bending of a beam, which adopts a parabolic profile and in turn modifies the electrostatic pressure. As demonstrated by Fig. 2b, this procedure successfully accounts for much of the curvature in Fig. 2a.

A striking feature in Fig. 2b is the vertical dark brown band at $n_2 \sim 0$, which, with increasing $E_2$, develops into a triangular green region. This indicates a very low conductance state at high electric field and charge neutrality point. Fig. 2c plots several line traces $G(n_2)$ at different $E_2$ values ranging from 0.21 to 0.73 V/nm. Each curve displays a minimum at $n_2 \approx 0$. For small values of $E_2$, the minimum is rather shallow, and the device's on/off ratio $\eta$, defined as the ratio between the maximum and minimum conductance values at a given $E_2$, is about 10. As $E_2$ increases, the device conductance at $n_2 \approx 0$ decreases dramatically, while that for large $n_2$ remains almost constant, resulting in a rapidly increasing $\eta$. For $E_2=0.73$ V/nm, $\eta>20,000$.

Fig. 2d plots the device resistance $R=1/G$ at $n_2=0$ vs. $E_2$ (top axis) and $V_2$ (bottom axis); the exponential increase in $R$ with increasing $V_2$ over nearly 3 decade suggests a field-induced opening of a band gap in the top gated region. In the simplest picture, transport could occur via thermally activated carriers. From tight binding calculations[2, 3, 9, 10], BLG's band structure adopts a "Mexican-hat" shape under an applied potential $V_2$, with a dispersion relation

$$E^{\pm-} \approx \pm \frac{eV_2}{2} \mp \frac{eV_2 v_F^2 \hbar^2 k^2}{t_\perp^2} \pm \frac{v_F^4 \hbar^4 k^4}{t_\perp^2 eV_2} \qquad (2)$$

which is valid for $v_F \hbar k \ll V \ll t_\perp$, and a band gap

$$\Delta = \frac{t_\perp eV_2}{\sqrt{t_\perp^2 + eV_2^2}} \qquad (3)$$

Here $t_\perp \sim$ 0.2-0.4 eV is the inter-layer hopping energy, and $k$ is the electron wave vector. From (3), $\Delta$ scales almost linearly with $V_2$ until it saturates at $t_\perp$; if screening is taken into account, its magnitude can be reduced by a factor of 2. In our devices, $\Delta \sim$ 0-0.1 eV[14]. If electrons are thermally activated to traverse the gapped region, we expect $R \sim exp(-eV_2/2k_BT)$, with an exponent $b \sim e/2k_BT \sim 19000$, where $k_B$ is Boltzmann's constant and $T$=0.3 K. However, from Fig. 2, the slope of the semi-log plot is ~36, or 2-3 orders of magnitude smaller than expected. This suggests that transport is across the top gated region is not thermally activated. Indeed, we observe little temperature dependence of conductance for $T$<1K.

Another transport mechanism is variable range hopping (VRH)[28], in which charge carriers are thermally activated to hop between localized states. For VRH in 2D, one expects $G \sim exp\left[-\left(T_0/T\right)^{1/3}\right]$ where $k_B T_0 = 4a\kappa^2/\rho_0$, $a$ is a dimensionless constant of order unity, $\kappa$ the coefficient of the exponential decay of the localized state, and $\rho_0$ is the density of states at the Fermi level[28]. Using the WKB approximation, we expect $\kappa \sim \sqrt{2m^* V_2}/\hbar$, where $m^*$ is the effective mass of charge carriers. From Eq. (2), $\rho_0 = \frac{2k}{\pi} \frac{\partial k}{\partial E}$ and $m^* = \hbar^2 / \frac{\partial^2 E}{\partial k^2}$, and ignoring the quadratic term in $v_F \hbar k / V_2$, we obtain $k_B T_0 \approx 4\pi a eV_2$.

To see if VRH can quantitatively account for the data, we fit the data in Fig. 2d to the expression $R = R_0 + A\exp\left[-\left(\frac{4\pi e}{k_B}aV_2\right)^{1/3}\right]$, where $R_0$ is the series resistance to account for the resistance of the non-top-gated region. Here we use $T_0/T=T_0$ because of our data's temperature independence below 1K. Satisfactory fit can be obtained by using the parameters $R_0$=5.3 k$\Omega$, $A$=1.85x10$^{-4}$ $\Omega$, and $a$=0.32, in agreement with the expectation that $a$ is a constant of order unity. This excellent agreement between VRH model and our data strongly suggests a successful band gap opening in BLG, and transport via variable range hopping between localized states that either lie within the gap[29-32] or are formed from disorder-induced charge puddles[33, 34].

We now focus on the device behavior in high magnetic fields. In the QH regime, the non-uniform charge density gives rise to regions of different filling factors, and, for bi-polar (i.e. *pnp* or *npn*) junctions, counter-propagating edge states. Consequently, the device conductance exhibit plateaus at fractional values of $e^2/h$ that arise from the mixing of edge states at the interfaces. For SLG, assuming full edge stage equilibration, the device conductance has been theoretically[16] and experimentally[16-18, 21] shown to obey the following formula,

$$G=e^2/h|v_2| \qquad \text{if } v_1v_2>0 \text{ and } |v_1| \geq |v_2|, \qquad (4a)$$

$$G = \frac{e^2}{h}\left(\frac{1}{|v_1|} \mp \frac{1}{|v_2|} + \frac{1}{|v_1|}\right)^{-1} \quad \begin{cases} -: \text{if } v_1v_2 > 0 \text{ and } |v_2|>|v_1| \\ +: \text{if } v_1v_2 < 0 \end{cases} \qquad (4b)$$

where $v_1, v_2$=..-6, -2, 2, 6, … are the filling factors in the areas outside and within the top-gated regions, respectively. For BLG, one expects these simple relations continue to hold, with $v_1, v_2$=..-8, -4, 4, 8… instead, yet they have never been experimentally verified.

Fig. 3a displays a typical data set $G(V_{tg}, V_{bg})$ measured at $B$=8T. The conductance map appears as a plaque of adjoined parallelograms, corresponding to different $v_1$ and $v_2$ combinations, similar to that observed in SLG devices. Fig. 3b plots a line trace $G(V_{tg})$ in units of $e^2/h$ at constant $v_1=4$. As $V_{tg}$ increases from -30 or $v_2=\sim-15$, $G$ decreases from 1.6, reaching a minimum of 1.35 at $v_2=-4$, then increases to a maximum plateau of 4 at $v_2=4$, before decreasing again. From Eq. (1), the conductance values are predicted to be 12/7, 8/5, 4/3 and 4, for $v_2=-12, -8, -4$ and 4, respectively, in good agreement with the data. Similarly, line traces at $v_1=8$ and 12 are shown in Fig. 2c, with conductance values reasonably accounted for by Eq. (4). We emphasize that this is the first time that the edge state equilibration is observed in bilayer *pnp* junctions, again underscoring the high quality of our junctions.

There is, however, one important discrepancy between Eq. (4) and our data – in Fig. 4c, $G$ reaches a minimum of $<10^{-7}$ S, or an insulating state, which cannot be obtained from (4) using $v_1, v_2=...-8, -4, 4, 8...$. This zero conductance state corresponds to $v_2=0$, indicating opening of a band gap in the top gated region and lifting of the orbital LL degeneracy. To investigate this insulating state further, we measure the device conductance as a function of $V_{tg}$ and $B$, at fixed $V_{bg}=19V$, as shown in Fig. 4a. Significant fluctuations are visible, with slopes in the $V_{tg}-B$ plane that are the same as those of the adjacent plateaus (Fig. 4d). These fluctuations arise from charging and localization from quantum dots in the bulk of the device, which consist of compressible regions that are surrounded by incompressible regions[21, 35]. Notably, fluctuations with zero slope have been observed, again indicating the presence of the $v_2=0$ state, which is not observed for monolayer graphene devices on substrate. These fluctuations also suggest significant spatial variations in density (up to $\pm v$) in the BLG sheet, hence confirming the presence of electron and hole puddles near the charge neutrality point, which could give rise to localized states, as discussed above.

We now focus on the insulating $v_2=0$ state. Fig. 4b plots $R(B)$ at $V_{tg}=-27V$ and $V_{bg}=19V$, or equivalently, $n_2=10^{11}$ cm$^{-2}$ and $E=0.24$ V/nm. At $B=0$, G~25 µS; with increasing $B$, $G$ decreases exponentially with an exponent $c \approx 0.70$, reaching 0.12µS at B=8T. Such exponential dependence of $G$ on $B$ has been observed in suspended bilayer devices[36], though with a much larger exponent, presumably because of the very low disorder of suspended devices.

Thus far we have observed two different insulating states in BLG. The first is induced by an external electric field at $B=0$, and the conductance decreases exponentially with $E_2$. The second insulating state occurs at finite electric *and* magnetic fields, and the conductance decreases exponentially with $B$. These two insulating states arise from the competition between the $E$ and $B$, which has been a topic of recent theoretical interest[6]. In principle, a transition between a layer polarized insulator and a QH ferromagnet insulator is expected to be observable in low-disorder devices, with a slope E/B~$10^7$ V/m/T. In our current device, any such transition is smeared by the relatively large number of disorder-induced impurity states. This would be an important direction for future studies of ultraclean double-gated BLG devices.

In conclusion, using bilayer graphene pnp junctions, we have demonstrated equilibration of QH edge states, insulating states induced by electric and/or magnetic fields, and transport via VRH in the gapped regime. In the future, experimental investigation of BLG with dual gates promise to yield a wealth of novel phenomena.

We thank Shaffique Adam for helpful discussions. The authors acknowledge the support of NSF CAREER DMR/0748910, ONR N00014-09-1-0724, GRC and the FENA Focus Center.

Note added in proof: After submission of this manuscript, we became aware of another experimental work on dual-gated bilayer graphene devices, Zou, K., Zhu, J. 2010, *Phys. Rev. B.* 82 (8), No. 081407(R)."

**Fig. 1. (a).** $G(V_{bg})$ for a bilayer graphene device at $T=260$mK and $B=0$. Inset: Device schematic.

**(b).** $G(V_{bg})$ of the device at $B=8$T. (c). LL fan diagram $G(V_{bg}, B)$ of the device.

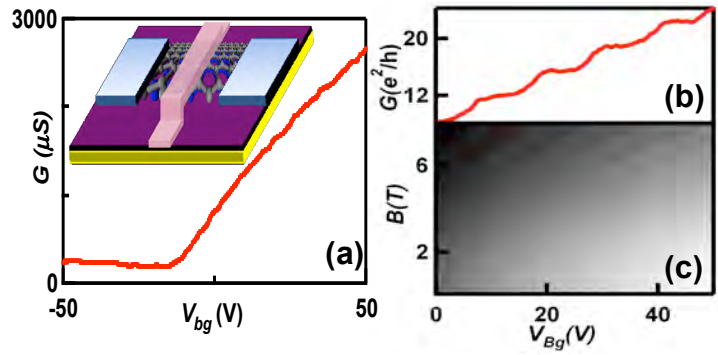

**Fig. 2. (a).** $G(V_{bg}, V_{tg})$ of the device at $B=0$. The junction polarities are indicated on the graph. **(b).** Data in (a) plotted in terms of $E_2$ and $n_2$. Note the logarithmic color scale. **(c).** (Top to bottom) Line traces $G(n_2)$ at $E_2=0.21, 0.49, 0.65$ and $0.73$ V/nm. **(d).** $R(E_2)$ at $n_2=0$. The line is a fit of the data to the VRH model, $R = R_0 + A\exp\left[-\left(\dfrac{4\pi e}{k_B}aV_2\right)^{1/3}\right]$, where $R_0$, $a$ and $A$ are fitting parameters.

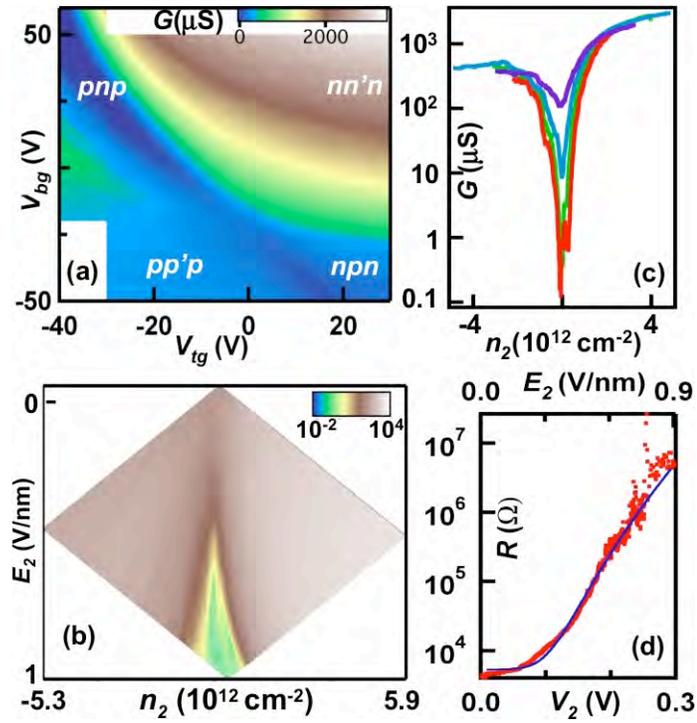

**Fig. 3. (a)** $G(V_{bg}, V_{tg})$ of the device at $B=8T$. **(b).** Line trace at $\nu_1=4$, *i.e.,* along the red dotted line in (a). **(c).** Line traces $G(V_{tg})$ at $\nu_1=8$ (yellow) and $\nu_1=12$ (purple).

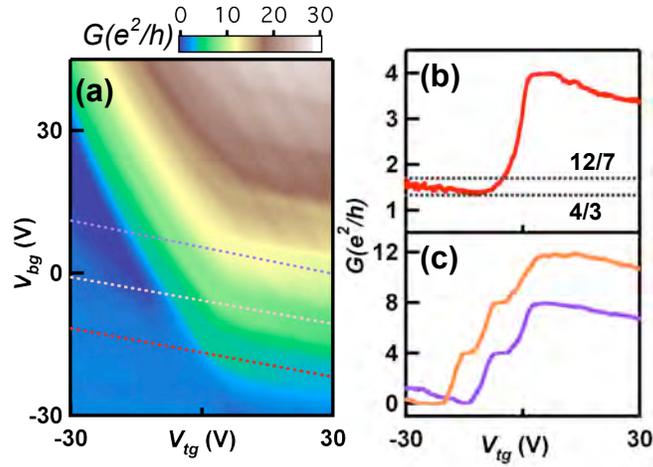

**Fig. 4. (a).** $G(V_{tg}, B)$ at $V_{bg}=19V$. **(b).** Data in (a) differentiated with respect to $V_{tg}$. **(c).** Line trace $G(B)$ taken at $V_{bg}=19V$ and $V_{tg}=-27V$. The blue line is a fit to an exponential function.

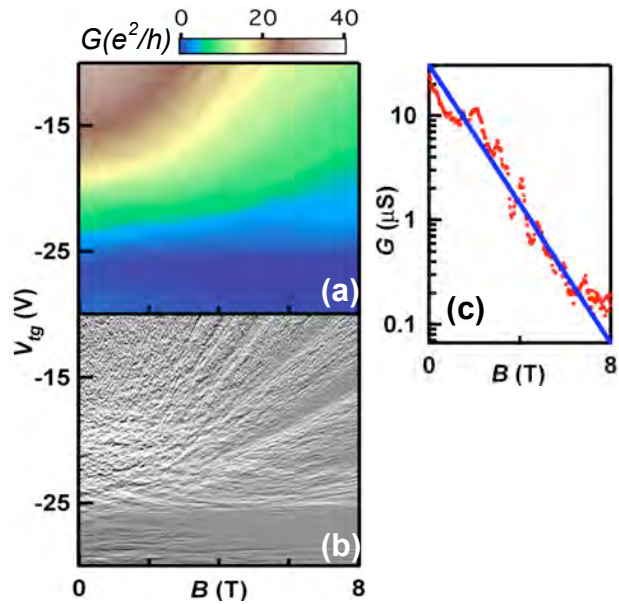